\documentclass[conference]{IEEEtran}

\usepackage{amsmath,amssymb,amsfonts,bm,mathtools}
\usepackage{cite}
\usepackage{microtype}

\usepackage{algorithm}
\usepackage{algpseudocode}

\begin{document}

\title{Joint Antenna Rotation and Polarforming for Satellite Communications}

\author{
    \IEEEauthorblockN{
    Xingxiang Peng\IEEEauthorrefmark{1},
    Qingqing Wu\IEEEauthorrefmark{1},
    Jingfeng Chen\IEEEauthorrefmark{1},
    Wei Feng\IEEEauthorrefmark{2},
    Wen Chen\IEEEauthorrefmark{1}, and
    Zan Li\IEEEauthorrefmark{3}
    }
    \IEEEauthorblockA{
        \IEEEauthorrefmark{1}School of Integrated Circuits, Shanghai Jiao Tong University, Shanghai 200240, China\\
        \IEEEauthorrefmark{2}Department of Electronic Engineering, Tsinghua University, Beijing 100084, China\\
        \IEEEauthorrefmark{3}School of Telecommunication Engineering, Xidian University, Xi'an 710071, China\\
        E-mails: peng\_xingxiang@sjtu.edu.cn, qingqingwu@sjtu.edu.cn, laowu3917@sjtu.edu.cn,\\
        fengwei@tsinghua.edu.cn, wenchen@sjtu.edu.cn, zanli@xidian.edu.cn
    }
}

\maketitle

\begin{abstract}
    This paper investigates joint antenna rotation and polarforming for low-Earth-orbit (LEO) communications, where a satellite equipped with rotatable dual-polarized antenna elements receives uplink signals from ground terminals (GTs) distributed across multiple clusters. In particular, antenna rotation redistributes directional gains across angularly separated clusters, while polarforming improves the separability of spatially correlated GT channels within each cluster. We formulate a weighted sum-rate maximization problem that jointly optimizes rotatable antenna (RA) orientation trajectories, element-wise polarformers, and digital receive filters under steering-range and inter-epoch slew constraints. To solve the resulting nonconvex problem, we develop a weighted minimum mean-square error (WMMSE)-based alternating algorithm, in which the optimal polarformer for each element is derived in semi-closed form, while an exact boresight parameterization yields convex feasible sets for RA-trajectory optimization, enabling Armijo-based gradient updates. Simulation results demonstrate that the proposed joint design outperforms benchmark schemes and reveal the complementary roles of antenna rotation in directional-gain allocation and polarforming in improving channel separability.
\end{abstract}


\section{Introduction}\label{sec:introduction}


Low-Earth-orbit (LEO) satellite networks are emerging as an important component of future sixth-generation (6G) non-terrestrial networks (NTNs), offering wide-area coverage and low-latency connectivity \cite{Giordani2021NTN}. Existing studies have investigated massive multiple-input multiple-output (MIMO) precoding and user grouping \cite{You2020LEOMIMO}, robust resource allocation under imperfect sensing \cite{Li2026LEODRL}, multi-objective resource management for LEO transmission \cite{Li2026SCMA}, and dynamic spectrum control for covert integrated air--ground communications \cite{Li2026CovertAirGround}.

Beyond conventional beamforming, reconfigurable antenna architectures provide additional degrees of freedom for shaping effective channels in LEO satellite communications. For example, movable antenna (MA) arrays have been investigated for dynamic satellite beam coverage and interference mitigation \cite{Zhu2025MASatellite}, as well as joint antenna positioning and beamforming at ground stations \cite{Wang2026MAGroundStation}. Unlike MAs, which change antenna positions to reconfigure array geometry, rotatable antennas (RAs) adjust element orientations to redistribute directional gains, thereby reshaping the relative strengths of signals arriving from different directions. RAs have been explored for multiuser transmission \cite{Peng1}, and interference management \cite{Peng2}. More recently, RA-enabled boresight alignment and beam tracking have been investigated for a single satellite-to-ground link \cite{Ma2026RASatellite}, while joint satellite selection and RA-boresight trajectory optimization have been studied for multi-satellite reception \cite{Peng2026RASatellite}.

However, MA- and RA-enabled spatial channel shaping alone may offer limited discrimination among users with closely aligned propagation directions, particularly in satellite communications, where long propagation distances lead to small angular separations and finite array apertures limit spatial resolution. Polarization provides an additional degree of freedom, as signals with similar spatial characteristics may still exhibit distinct polarization responses. By adaptively combining orthogonal polarization components, polarforming enables polarization-domain channel shaping. The principles and benefits of polarforming have been discussed in \cite{Ding2026Polarforming}, while its joint design with antenna positioning has been investigated in MA systems \cite{Zhou2026MAPolarforming}. Polarforming has also been integrated with antenna positioning and rotation for wireless sensing and communication \cite{Shao2026Polarforming}. However, these studies have mainly focused on terrestrial systems, while the potential of combining spatial and polarization-domain channel shaping in satellite systems merits further investigation. In particular, the dynamic satellite geometry continuously changes propagation directions and polarization projections, complicating the joint exploitation of antenna directivity and polarization diversity.

In this paper, we investigate a polarization-aware RA architecture for LEO satellite communications, where ground terminals (GTs) are distributed across geographically separated clusters. From the satellite's perspective, different clusters are separated in angle, whereas GTs within each cluster have similar propagation directions and may thus exhibit highly correlated spatial channels. By reconfiguring antenna orientations, the satellite can redistribute directional gains across clusters, while element-wise polarforming exploits polarization diversity to improve intra-cluster channel separability. We formulate a weighted sum-rate maximization problem that jointly optimizes RA orientation trajectories, element-wise polarformers, and digital receive filters under steering-range and inter-epoch slew constraints. To tackle the resulting nonconvex problem, we develop a weighted minimum mean-square error (WMMSE)-based alternating algorithm, in which the optimal polarformer for each element is derived in semi-closed form, while RA trajectories are optimized via Armijo-based gradient updates. Numerical results demonstrate the benefits of jointly exploiting directional and polarization-domain channel shaping in multiuser satellite communications.

\section{System Model and Problem Formulation}
\label{sec:system_model}

\begin{figure}[t]
    \centering
    \includegraphics[width=\columnwidth]{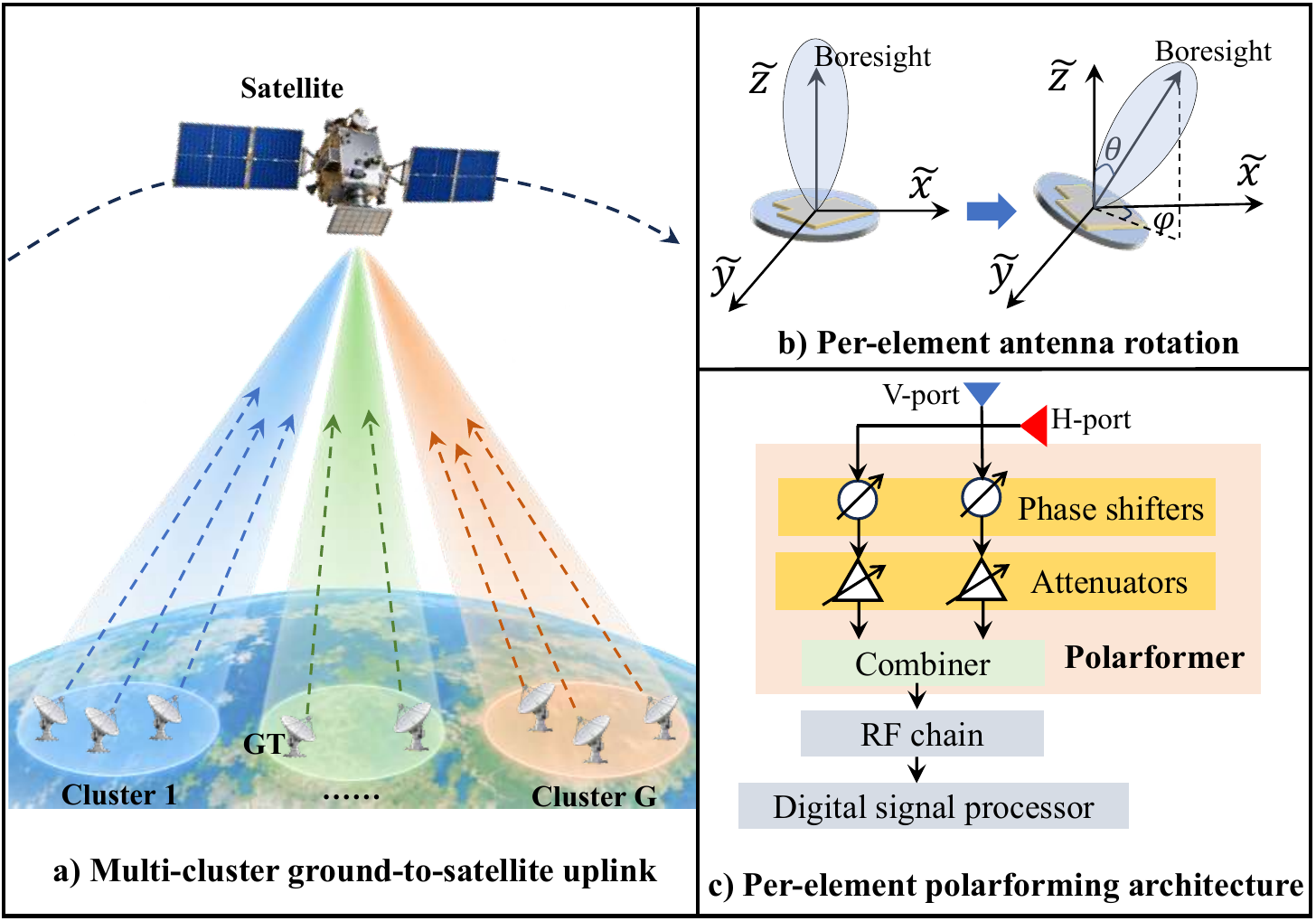}
    \caption{Multi-cluster ground-to-satellite communications, where a satellite equipped with rotatable dual-polarized antenna elements receives uplink signals.}
    \label{fig:sys_mod}
\end{figure}

As illustrated in Fig.~\ref{fig:sys_mod}, we consider a LEO satellite receiving independent data streams from $K$ GTs over a shared time--frequency resource. The GTs form $G$ geographical clusters, with $\mathcal{K}=\bigcup_{g=1}^{G}\mathcal{K}_g$ and $K=\sum_{g=1}^{G}|\mathcal{K}_g|$. The satellite carries a nadir-facing $M_x\times M_y$ uniform planar array (UPA) of $M=M_xM_y$ RA elements supporting independent boresight steering and receive polarforming. Each GT uses a single-port linearly polarized antenna with unit isotropic gain.

\subsection{Two-Timescale Operation Protocol}
\label{subsec:protocol}

To account for the slower mechanical reconfiguration of the RAs relative to electronic polarforming and digital combining, we adopt a two-timescale protocol. Specifically, the RA boresights are held fixed within each control epoch, whereas polarforming and digital combining are updated per slot. The observation interval contains $N$ data-bearing slots of duration $\Delta t$ and is divided into $L$ epochs with $Q=N/L$ slots each, where $\mathcal{N}_{\ell}=\{(\ell-1)Q+1,\ldots,\ell Q\}$. Each epoch $\ell\geq2$ is preceded by a guard interval $T_{\rm g}$ for mechanical slewing, yielding an observation duration of $T_{\rm ob}=N\Delta t+(L-1)T_{\rm g}$. The midpoint of slot $n$ is
\begin{equation}
    t_n=\left(n-\frac{1}{2}\right)\Delta t+\left(\left\lceil\frac{n}{Q}\right\rceil-1\right)T_{\rm g}.
    \label{eq:slot_midpoint}
\end{equation}
The geometry and corresponding channels are evaluated at $t_n$.

\subsection{Dynamic Satellite and Clustered-GT Geometry}
\label{subsec:geometry}

The Earth is modeled as a sphere of radius $R_{\mathrm{E}}$, and the satellite follows a circular orbit of altitude $H_{\mathrm{s}}$ and inclination $\beta$. Following \cite{Peng2026RASatellite}, let $\bm{r}_{\mathrm{s}}^{\mathrm{E}}(t)$ denote its ECEF position and $\bm{r}_k^{\mathrm{E}}$ the fixed ECEF position of GT $k$. We adopt a satellite-centric Cartesian coordinate system (SCCS) whose $x$-, $y$-, and $z$-axes point along the along-track direction, opposite to the orbital normal, and toward the geocenter, respectively. Let $\bm{T}_{\mathrm{s}}(t)$ denote the SCCS-to-ECEF rotation matrix. The slant range and satellite-to-GT look directions are
\begin{subequations}
\label{eq:link_geometry}
    \begin{align}
        d_k[n]&=\|\bm{r}_k^{\mathrm{E}}-\bm{r}_{\mathrm{s}}^{\mathrm{E}}(t_n)\|_2,\label{eq:slant_range}\\
        \bm{d}_k^{\mathrm{E}}[n]&=\frac{\bm{r}_k^{\mathrm{E}}-\bm{r}_{\mathrm{s}}^{\mathrm{E}}(t_n)}{d_k[n]},\\
        \bm{d}_k^{\mathrm{S}}[n]&=\bm{T}_{\mathrm{s}}^T(t_n)\bm{d}_k^{\mathrm{E}}[n].\label{eq:look_directions}
    \end{align}
\end{subequations}
The uplink propagation direction is $\bm{u}_k^{\mathrm{E}}[n]=-\bm{d}_k^{\mathrm{E}}[n]$. Let $\bm{E}_k\triangleq[\bm{e}_{E,k},\bm{e}_{N,k}]$ collect the local east and north unit vectors in ECEF, and let $\bm{e}_{U,k}=\bm{r}_k^{\mathrm{E}}/R_{\mathrm{E}}$ denote the local up direction. We consider a common service interval satisfying $\epsilon_k[n]=\arcsin(\bm{e}_{U,k}^T\bm{u}_k^{\mathrm{E}}[n])\geq\epsilon_{\min}$ for all $k,n$

\subsection{Satellite-Side RA and Polarforming Configuration}
\label{subsec:ra_polarforming}

Let $\bm{e}_x$, $\bm{e}_y$, and $\bm{e}_z$ denote the canonical SCCS basis. The fixed UPA lies in the $x$--$y$ plane, with element positions
\begin{equation}
    \bm{p}_m=\begin{bmatrix}\left(m_x-\frac{M_x-1}{2}\right)d_x & \left(m_y-\frac{M_y-1}{2}\right)d_y & 0\end{bmatrix}^T,
    \label{eq:upa_element_position}
\end{equation}
where $m=m_x+m_yM_x+1$, $m_x\in\{0,\ldots,M_x-1\}$, $m_y\in\{0,\ldots,M_y-1\}$, and $d_x,d_y$ are the element spacings. The unit boresight $\bm{f}_m[\ell]$ has a maximum tilt $\theta_{\max}\in[0,\pi/2)$ from $\bm{e}_z$. Its inter-epoch rotation is limited to $\Omega_{\mathrm{g}}=\omega_{\max}T_{\mathrm{g}}$, where $\omega_{\max}$ is the maximum mechanical angular speed.

Each element has two co-located orthogonal polarization ports with the SCCS reference basis
\begin{subequations}
\label{eq:local_polarization_basis}
    \begin{align}
        \bm{b}_{m,H}[\ell]&=\frac{(\bm{I}_3-\bm{f}_m[\ell]\bm{f}_m^T[\ell])\bm{e}_x}{\|(\bm{I}_3-\bm{f}_m[\ell]\bm{f}_m^T[\ell])\bm{e}_x\|_2},\label{eq:local_h_axis}\\
        \bm{b}_{m,V}[\ell]&=\bm{f}_m[\ell]\times\bm{b}_{m,H}[\ell].\label{eq:local_v_axis}
    \end{align}
\end{subequations}
This basis is well defined for $\theta_{\max}<\pi/2$. For $n\in\mathcal{N}_{\ell}$, its ECEF representation is
\begin{equation}
    \bm{B}_m[\ell,n]=\bm{T}_{\mathrm{s}}(t_n)[\bm{b}_{m,H}[\ell],\bm{b}_{m,V}[\ell]].
    \label{eq:satellite_port_basis}
\end{equation}
The two polarization-port signals are coherently combined by a unit-norm polarformer $\bm{c}_m[n]\in\mathbb{C}^2$.

\subsection{Polarization-Aware Uplink Channel Model}
\label{subsec:channel_model}

The GT--satellite links are modeled as far-field, polarization-preserving line-of-sight (LoS) channels. For $n\in\mathcal{N}_{\ell}$, the directional receive gain and free-space path gain are
\begin{align}
    G_{m,k}[n]&=\kappa_{\max}[\bm{f}_m^T[\ell]\bm{d}_k^{\mathrm{S}}[n]]_+^{2p},\label{eq:directional_gain}\\
    \beta_k[n]&=\left(\frac{\lambda_{\mathrm{c}}}{4\pi d_k[n]}\right)^2,\label{eq:path_loss}
\end{align}
where $[x]_+\triangleq\max\{x,0\}$, $p\geq0$ controls the antenna directivity, $\kappa_{\max}=2(2p+1)$, and $\lambda_{\mathrm{c}}$ is the carrier wavelength. With respect to the local east--north basis \(\bm E_k\), the GT polarization is represented by the Jones vector $\bm{t}_k=[\cos\xi_k,\sin\xi_k]^T$, where $\xi_k\sim\mathcal{U}[0,\pi)$ is independently generated and remains fixed over the observation interval. To characterize polarization coupling, let $\bm{Z}_k[n]\in\mathbb{R}^{3\times2}$ denote an orthonormal basis of the transverse wavefront plane in ECEF, satisfying $\bm{Z}_k^T[n]\bm{Z}_k[n]=\bm{I}_2$ and $\bm{Z}_k^T[n]\bm{u}_k^{\mathrm{E}}[n]=\bm{0}$. Using this basis, the transmit- and receive-side polarization projections are given by
\begin{subequations}
    \begin{align}
    \bm{P}_k[n]&=\bm{Z}_k^T[n]\bm{E}_k,\\
    \bm{Q}_{m,k}[\ell,n]&=\bm{B}_m^T[\ell,n]\bm{Z}_k[n],
    \label{eq:polarization_projections}
    \end{align}
\end{subequations}
and the resulting polarization response is given by \cite{Shao2026Polarforming}
\begin{equation}
    \eta_{m,k}[\ell,n]=\bm{c}_m^H[n]\bm{Q}_{m,k}[\ell,n]\bm{P}_k[n]\bm{t}_k.
    \label{eq:polarization_response}
\end{equation}
This response is independent of the choice of $\bm{Z}_k[n]$, since $\bm{Z}_k[n]\bm{Z}_k^T[n]=\bm{I}_3-\bm{u}_k^{\mathrm{E}}[n](\bm{u}_k^{\mathrm{E}}[n])^T$.

Combining the propagation loss, directional gain, polarization response, and array-induced phase, the effective channel from GT $k$ to satellite element $m$ in slot $n\in\mathcal{N}_{\ell}$ is
\begin{equation}
\begin{aligned}
    h_{m,k}[n]={}&\sqrt{\beta_k[n]G_{m,k}[n]}\,\eta_{m,k}[\ell,n]e^{-j\frac{2\pi}{\lambda_{\rm c}}d_k[n]} e^{j\frac{2\pi}{\lambda_{\rm c}}\bm{p}_m^{T}\bm{d}_k^{\rm S}[n]}.
\end{aligned}
    \label{eq:per_element_channel}
\end{equation}
Stacking the element-wise channel coefficients, we obtain the uplink channel vector of GT $k$ as
\begin{equation}
    \bm{h}_k[n]=\begin{bmatrix}h_{1,k}[n] & \cdots & h_{M,k}[n]\end{bmatrix}^{T}.
    \label{eq:channel_vector}
\end{equation}
The resulting uplink channel captures the joint effects of satellite geometry, RA boresight steering, and receive polarforming.

\subsection{Multiuser Uplink Reception and Problem Formulation}
\label{subsec:signal_model}

Let $P_k$ denote the fixed transmit power of GT $k$ and $s_k[n]\sim\mathcal{CN}(0,1)$ its unit-power symbol. The received signal is
\begin{equation}
    \bm{y}[n]=\sum_{k=1}^{K}\sqrt{P_k}\bm{h}_k[n]s_k[n]+\bm{z}[n],
    \label{eq:received_signal}
\end{equation}
where $\bm{z}[n]\sim\mathcal{CN}(\bm{0},\sigma^2\bm{I}_M)$. Let $\bm{w}_k[n]\in\mathbb{C}^M$ denote a nonzero linear receive filter for GT $k$. The corresponding symbol estimate is $\widehat{s}_k[n]=\bm{w}_k^H[n]\bm{y}[n]$, yielding the signal-to-interference-plus-noise ratio (SINR)
\begin{equation}
    \gamma_k[n]=\frac{P_k|\bm{w}_k^H[n]\bm{h}_k[n]|^2}{\sum_{j\neq k}P_j|\bm{w}_k^H[n]\bm{h}_j[n]|^2+\sigma^2\|\bm{w}_k[n]\|_2^2}.
    \label{eq:sinr}
\end{equation}
For bandwidth $B$, the achievable rate is $R_k[n]=B\log_2(1+\gamma_k[n])$. With GT weights $\mu_k>0$ and guard overhead included, the weighted effective sum rate is given by
\begin{equation}
    \overline{R}_{\mathrm{eff}}=\frac{\Delta t}{T_{\mathrm{ob}}}\sum_{\ell=1}^{L}\sum_{n\in\mathcal{N}_{\ell}}\sum_{k=1}^{K}\mu_k R_k[n].
    \label{eq:effective_wsr}
\end{equation}
Define $\mathcal{F}\triangleq\{\bm{f}_m[\ell]\}_{m,\ell}$, $\mathcal{C}\triangleq\{\bm{c}_m[n]\}_{m,n}$, and $\mathcal{W}\triangleq\{\bm{w}_k[n]\}_{k,n}$. We formulate the joint optimization problem as
\begin{subequations}
\label{prob:joint_design}
    \begin{align}
        \underset{\mathcal{F},\mathcal{C},\mathcal{W}}{\operatorname{max}}\quad &\overline{R}_{\mathrm{eff}}\label{prob:objective}\\
        \operatorname{s.t.}\quad &\|\bm{f}_m[\ell]\|_2=1,\quad \forall m,\ell,\label{prob:unit_boresight}\\
        &\bm{f}_m^T[\ell]\bm{e}_z\geq\cos\theta_{\max},\quad \forall m,\ell,\label{prob:boresight_constraint}\\
        &\bm{f}_m^T[\ell-1]\bm{f}_m[\ell]\geq\cos\Omega_{\mathrm{g}},\quad \forall m,\ \ell\geq2,\label{prob:slew_constraint}\\
        &\|\bm{c}_m[n]\|_2=1,\quad \forall m,n.\label{prob:polarforming_constraint}
    \end{align}
\end{subequations}
Constraints \eqref{prob:unit_boresight}--\eqref{prob:slew_constraint} enforce the steering range and inter-epoch slew limits of the RA boresights, while \eqref{prob:polarforming_constraint} imposes the unit-norm constraint on the polarformers. Problem \eqref{prob:joint_design} is highly nonconvex due to the nonlinear coupling among the RA boresights, polarformers, and receive filters through the multiuser channels, together with the unit-norm constraints and the inter-epoch coupling introduced by the slew limits.

\section{The Proposed Solution}
\label{sec:proposed_solution}

To solve \eqref{prob:joint_design}, we reformulate the weighted sum-rate maximization into an equivalent WMMSE problem, enabling block-wise optimization of the receive filters, polarformers, and RA boresights. We then derive efficient polarforming and boresight-trajectory updates, and integrate them into an alternating framework that ensures a nonincreasing and convergent WMMSE objective sequence.

\subsection{WMMSE Reformulation}
\label{subsec:sol_wmmse}

We first transform the weighted sum-rate maximization into an equivalent WMMSE problem following the standard rate--MSE relationship \cite{Shi2011WMMSE}. For GT $k$ in slot $n$, the symbol estimation MSE associated with the linear receive filter $\bm{w}_k[n]$ is defined as
\begin{align}
    e_k[n] &\triangleq \mathbb{E}\left\{\left|\bm{w}_k^H[n]\bm{y}[n]-s_k[n]\right|^2\right\} \nonumber\\
    &=\left|\sqrt{P_k}\bm{w}_k^H[n]\bm{h}_k[n]-1\right|^2 \nonumber\\
    &\quad+\sum_{j\neq k}P_j\left|\bm{w}_k^H[n]\bm{h}_j[n]\right|^2+\sigma^2\|\bm{w}_k[n]\|_2^2.
    \label{eq:sol_mse}
\end{align}
Introducing a positive auxiliary MSE weight $\nu_k[n]>0$ for each GT and slot, the weighted sum-rate maximization in \eqref{prob:joint_design} can be equivalently reformulated as
\begin{subequations}
\label{prob:sol_wmmse}
    \begin{align}
        \underset{\mathcal{F},\mathcal{C},\mathcal{W},\boldsymbol{\nu}}{\operatorname{min}}\quad &\sum_{n=1}^{N}\sum_{k=1}^{K}\mu_k\left(\nu_k[n]e_k[n]-\ln\nu_k[n]-1\right)
        \label{prob:sol_wmmse_obj}\\
        \operatorname{s.t.}\quad &\|\bm{f}_m[\ell]\|_2=1,\qquad \forall m,\ell,
        \label{prob:sol_wmmse_unit_boresight}\\
        &\bm{f}_m^T[\ell]\bm{e}_z\geq\cos\theta_{\max},\qquad \forall m,\ell,
        \label{prob:sol_wmmse_boresight}\\
        &\bm{f}_m^T[\ell-1]\bm{f}_m[\ell]\geq\cos\Omega_{\rm g},\quad \forall m,\ \ell\geq2,
        \label{prob:sol_wmmse_slew}\\
        &\|\bm{c}_m[n]\|_2=1,\qquad \forall m,n,
        \label{prob:sol_wmmse_polarforming}\\
        &\nu_k[n]>0,\qquad \forall k,n,
    \label{prob:sol_wmmse_weight}
    \end{align}
\end{subequations}
where $\boldsymbol{\nu}\triangleq\{\nu_k[n]\}_{k,n}$. The reformulation exposes a block structure that decomposes the coupled problem into tractable subproblems. We therefore adopt an alternating framework to iteratively optimize the receive filters and MSE weights, polarformers, and RA boresights. For fixed $(\mathcal{F},\mathcal{C})$, the optimal receive filters and MSE weights admit closed-form updates as
\begin{align}
\bm{w}_k^\star[n]&=\bigg(\sum_{j=1}^{K}P_j\bm{h}_j[n]\bm{h}_j^H[n]+\sigma^2\bm{I}_M\bigg)^{-1}\sqrt{P_k}\bm{h}_k[n],\label{eq:sol_w_update}\\
\nu_k^\star[n]&=\frac{1}{e_k^\star[n]},
\label{eq:sol_nu_update}
\end{align}
where $e_k^\star[n]$ denotes the MSE in \eqref{eq:sol_mse} evaluated at $\bm{w}_k^\star[n]$.

\subsection{Polarforming Update}
\label{subsec:sol_polarforming}

With given $(\mathcal{F},\mathcal{W},\boldsymbol{\nu})$, the polarforming optimization is decoupled across slots, since $\{\bm{c}_m[n]\}_{m=1}^{M}$ affects only the channel coefficients and MSE terms in slot $n$. As all resulting subproblems share the same structure, we omit the slot index and consider an arbitrary slot, for which the polarforming problem reduces to
\begin{subequations}    \label{prob:sol_polarforming}
    \begin{align}
    \underset{\{\bm{c}_m\}_{m=1}^{M}}{\operatorname{min}}\quad &\sum_{k=1}^{K}\mu_k\nu_k e_k\\
    \operatorname{s.t.}\quad &\|\bm{c}_m\|_2=1,\qquad \forall m. 
    \end{align}
\end{subequations}
For convenience, we rewrite the channel in \eqref{eq:per_element_channel} by explicitly separating the dependence on $\bm{c}_m$. Specifically, we define
\begin{equation}
    \bm{a}_{m,k}\triangleq\sqrt{P_k\beta_k G_{m,k}}\,\bm{Q}_{m,k}\bm{P}_k\bm{t}_k e^{-j\frac{2\pi}{\lambda_{\rm c}}d_k}e^{j\frac{2\pi}{\lambda_{\rm c}}\bm{p}_m^T\bm{d}_k^{\rm S}},
    \label{eq:sol_two_port_channel}
\end{equation}
such that $\sqrt{P_k}h_{m,k}=\bm{c}_m^H\bm{a}_{m,k}$. Let $w_{k,m}$ denote the $m$th entry of $\bm{w}_k$. Substituting the above relation into \eqref{eq:sol_mse}, the symbol estimation MSE for GT $k$ can be expressed as
\begin{equation}
    e_k=\sum_{j=1}^{K}\left|\sum_{m=1}^{M}w_{k,m}^*\bm{c}_m^H\bm{a}_{m,j}-\delta_{kj}\right|^2+\sigma^2\|\bm{w}_k\|_2^2,
    \label{eq:sol_mse_polarforming}
\end{equation}
where $\delta_{kj}$ is the Kronecker delta. Equation \eqref{eq:sol_mse_polarforming} shows that the polarformers of different antenna elements remain coupled in the considered slot. We therefore update them cyclically, with each $\bm{c}_m$ optimized while the other polarformers are fixed. Specifically, for the update of element $m$, define
\begin{equation}
    \rho_{k,j}^{(-m)}\triangleq\sum_{m'\neq m}w_{k,m'}^*\bm{c}_{m'}^H\bm{a}_{m',j}-\delta_{kj}.
    \label{eq:sol_polar_residual}
\end{equation}
Then \eqref{eq:sol_mse_polarforming} can be further expressed as
\begin{equation}
    e_k=\sum_{j=1}^{K}\left|\rho_{k,j}^{(-m)}+w_{k,m}^*\bm{c}_m^H\bm{a}_{m,j}\right|^2+\sigma^2\|\bm{w}_k\|_2^2.
    \label{eq:sol_mse_element}
\end{equation}
Since the noise term is independent of $\bm{c}_m$, the element-wise polarforming subproblem is
\begin{equation}
    \underset{\|\bm{c}_m\|_2=1}{\operatorname{min}}\quad \sum_{k=1}^{K}\sum_{j=1}^{K}\mu_k\nu_k\left|\rho_{k,j}^{(-m)}+w_{k,m}^*\bm{c}_m^H\bm{a}_{m,j}\right|^2.
    \label{prob:sol_element_polarforming_raw}
\end{equation}
Expanding the quadratic terms and dropping the constants independent of $\bm{c}_m$, \eqref{prob:sol_element_polarforming_raw} reduces to
\begin{equation}
    \underset{\|\bm{c}_m\|_2=1}{\operatorname{min}}\quad \bm{c}_m^H\bm{Q}_m\bm{c}_m-2\operatorname{Re}\{\bm{c}_m^H\bm{b}_m\},
    \label{prob:sol_element_polarforming}
\end{equation}
where $\bm{Q}_m \triangleq\sum_{k=1}^{K}\sum_{j=1}^{K}\mu_k\nu_k|w_{k,m}|^2\bm{a}_{m,j}\bm{a}_{m,j}^H$, and $\bm{b}_m \triangleq -\sum_{k=1}^{K}\sum_{j=1}^{K}\mu_k\nu_k\big(\rho_{k,j}^{(-m)}\big)^*w_{k,m}^*\bm{a}_{m,j}$. By the Lagrangian optimality conditions, the optimal solution of \eqref{prob:sol_element_polarforming} satisfies
\begin{subequations}
\label{eq:sol_polar_kkt}
    \begin{align}
    (\bm{Q}_m+\lambda_m^\star\bm{I}_2)\bm{c}_m^\star &=\bm{b}_m,
    \label{eq:sol_polar_kkt_stationarity}\\
    \|\bm{c}_m^\star\|_2 &=1.
    \label{eq:sol_polar_kkt_norm}
    \end{align}
\end{subequations}
where $\lambda_m^\star$ denotes the Lagrange multiplier for the unit-norm constraint, with $\bm{Q}_m+\lambda_m^\star\bm{I}_2\succeq\bm{0}$.
The optimal $\bm{c}_m^\star$ and $\lambda_m^\star$ can be obtained by solving the coupled equations \eqref{eq:sol_polar_kkt_stationarity} and \eqref{eq:sol_polar_kkt_norm}. Specifically, for a given $\lambda_m$, \eqref{eq:sol_polar_kkt_stationarity} provides a parameterized solution for $\bm{c}_m$, while \eqref{eq:sol_polar_kkt_norm} is used to determine the optimal $\lambda_m^\star$. To efficiently solve the above conditions, we perform the eigen-decomposition
\begin{equation}
    \bm{Q}_m=\bm{U}_m\operatorname{diag}(q_{m,1},q_{m,2})\bm{U}_m^H,
    \label{eq:sol_polar_eigendecomposition}
\end{equation}
where $q_{m,1}\leq q_{m,2}$. If $q_{m,1}=q_{m,2}$, the optimal solution is $\bm{c}_m^\star=\bm{b}_m/\|\bm{b}_m\|_2$ for $\bm{b}_m\neq\bm{0}$, while any unit-norm vector is optimal otherwise. For $q_{m,1}<q_{m,2}$, let $\Delta_m=q_{m,2}-q_{m,1}$. We first check whether $[\bm{U}_m^H\bm{b}_m]_1=0$ and $|[\bm{U}_m^H\bm{b}_m]_2|\leq\Delta_m$. If both conditions hold, the optimal multiplier is $\lambda_m^\star=-q_{m,1}$, and a globally optimal polarformer is then given by
\begin{equation}
    \bm{c}_m^\star=\bm{U}_m
    \begin{bmatrix}
        \sqrt{1-\left|[\bm{U}_m^H\bm{b}_m]_2/\Delta_m\right|^2}\\
        [\bm{U}_m^H\bm{b}_m]_2/\Delta_m
    \end{bmatrix}.
    \label{eq:sol_polar_singular}
\end{equation}
Otherwise, $\lambda_m^\star>-q_{m,1}$, and \eqref{eq:sol_polar_kkt_stationarity} yields
\begin{equation}
    \bm{c}_m(\lambda_m)=\bm{U}_m\operatorname{diag}\!\bigg(\frac{1}{q_{m,1}+\lambda_m},\frac{1}{q_{m,2}+\lambda_m}\bigg)\bm{U}_m^H\bm{b}_m.
    \label{eq:sol_polar_parametric}
\end{equation}
Enforcing the unit-norm constraint gives
\begin{equation}
    \sum_{i=1}^{2}\frac{|[\bm{U}_m^H\bm{b}_m]_i|^2}{(q_{m,i}+\lambda_m^\star)^2}=1.
    \label{eq:sol_polar_secular}
\end{equation}
The optimal multiplier $\lambda_m^\star$ is then determined through a one-dimensional search, yielding the optimal polarformer $\bm{c}_m^\star$.

\subsection{Boresight-Trajectory Update}
\label{subsec:sol_trajectory}

With given $(\mathcal{C},\mathcal{W},\boldsymbol{\nu})$, the optimization for the RA boresights reduces to
\begin{subequations}
    \label{prob:sol_boresight}
    \begin{align}
    \underset{\mathcal{F}}{\operatorname{min}}\quad &\sum_{n=1}^{N}\sum_{k=1}^{K}\mu_k\nu_k[n]e_k[n]
    \label{prob:sol_boresight_obj}\\
    \operatorname{s.t.}\quad &\|\bm{f}_m[\ell]\|_2=1,\qquad \forall m,\ell,
    \label{prob:sol_boresight_unit}\\
    &\bm{f}_m^T[\ell]\bm{e}_z\geq\cos\theta_{\max},\qquad \forall m,\ell,
    \label{prob:sol_boresight_cap}\\
    &\bm{f}_m^T[\ell-1]\bm{f}_m[\ell]\geq\cos\Omega_{\rm g},\qquad \forall m,\ \ell\geq2.
    \label{prob:sol_boresight_slew}
\end{align}
\end{subequations}
Unlike the slot-wise polarforming variables, the boresights of adjacent epochs are coupled through the slew constraints. To facilitate the boresight-trajectory optimization while preserving feasibility, we first introduce an exact two-dimensional parameterization of each boresight. Since $\theta_{\max}<\pi/2$, any feasible $\bm{f}_m[\ell]$ can be uniquely represented as
\begin{equation}
    \bm{f}_m[\ell]=\bm{f}(\bm{x}_m[\ell])\triangleq\begin{bmatrix}\bm{x}_m[\ell]\\ \sqrt{1-\|\bm{x}_m[\ell]\|_2^2}\end{bmatrix},
    \label{eq:sol_cap_parameterization}
\end{equation}
where $\bm{x}_m[\ell]\in\mathbb{R}^2$ collects the first two SCCS components of $\bm{f}_m[\ell]$ and satisfies
\begin{equation}
    \|\bm{x}_m[\ell]\|_2\leq\sin\theta_{\max}.
    \label{eq:sol_cap_constraint}
\end{equation}
Thus, the unit-norm and steering-range constraints in \eqref{prob:sol_boresight_unit} and \eqref{prob:sol_boresight_cap} are automatically satisfied. Let $\bm{X}\triangleq\{\bm{x}_m[\ell]\}_{m,\ell}$ and $c_{\rm g}\triangleq\cos\Omega_{\rm g}$. The slew constraint \eqref{prob:sol_boresight_slew} can be written as
\begin{equation}
    g_{m,\ell}(\bm{X})\triangleq c_{\rm g}-\bm{f}^T(\bm{x}_m[\ell-1])\bm{f}(\bm{x}_m[\ell])\leq0.
    \label{eq:sol_slew_constraint}
\end{equation}
Substituting \eqref{eq:sol_cap_parameterization} into this slew constraint gives
\begin{align}
    &g_{m,\ell}(\bm{X})
    \triangleq{}c_{\rm g}
    -\bm{x}_m^T[\ell-1]\bm{x}_m[\ell] \nonumber\\
    &\qquad -\sqrt{1-\|\bm{x}_m[\ell-1]\|_2^2}
    \sqrt{1-\|\bm{x}_m[\ell]\|_2^2}
    \leq0.
    \label{eq:sol_slew_constraint1}
\end{align}
Since $\theta_{\max}<\pi/2$, the constraint in \eqref{eq:sol_cap_constraint} ensures $\|\bm{x}_m[\ell]\|_2<1$ for all $m,\ell$. Over this domain, $g_{m,\ell}(\bm{X})$ is jointly convex in $\bm{x}_m[\ell-1]$ and $\bm{x}_m[\ell]$, as can be verified from its positive-semidefinite Hessian. Hence, the boresight steering and slew constraints define a convex feasible set in $\bm{X}$. Together with \eqref{eq:sol_cap_constraint}, it yields the following reformulation of the boresight-trajectory problem:
\begin{subequations}
\label{prob:sol_boresight_approx}
\begin{align}
    \underset{\bm{X}}{\operatorname{min}}\quad
    &J(\bm{X})\triangleq\sum_{n=1}^{N}\sum_{k=1}^{K}\mu_k\nu_k[n]e_k[n]
    \label{prob:sol_boresight_approx_obj}\\
    \operatorname{s.t.}\quad
    &\|\bm{x}_m[\ell]\|_2\leq\sin\theta_{\max},\qquad \forall m,\ell,
    \label{prob:sol_boresight_approx_cap}\\
    &g_{m,\ell}(\bm{X})\leq0,\qquad \forall m,\ \ell\geq2.
    \label{prob:sol_boresight_approx_slew}
\end{align}
\end{subequations}
Although the constraints in \eqref{prob:sol_boresight_approx} are convex, the objective $J(\bm{X})$ remains nonconvex due to the nonlinear dependence of the channels on the RA boresights. We therefore adopt a feasible first-order update that combines objective linearization with an Armijo line search. Specifically, we linearize $J(\bm{X})$ at the current trajectory $\bm{X}^r$. Let
\begin{equation}
    \bm{G}^r\triangleq\nabla_{\bm{X}}J(\bm{X}^r).
    \label{eq:sol_boresight_gradient}
\end{equation}
Dropping the constant term $J(\bm{X}^r)$, the resulting convex subproblem is given by
\begin{subequations}
    \label{prob:sol_trajectory_oracle}
    \begin{align}
    \underset{\bm{X}}{\operatorname{min}}\quad
    &\left\langle\bm{G}^r,\bm{X}-\bm{X}^r\right\rangle
    \label{prob:sol_trajectory_oracle_obj}\\
    \operatorname{s.t.}\quad
    &\|\bm{x}_m[\ell]\|_2\leq\sin\theta_{\max},\qquad \forall m,\ell,
    \label{prob:sol_trajectory_oracle_cap}\\
    &g_{m,\ell}(\bm{X})\leq0,\qquad \forall m,\ \ell\geq2.
    \label{prob:sol_trajectory_oracle_slew}
    \end{align}
\end{subequations}
Since both the linearized objective and the constraints are separable across antenna elements, \eqref{prob:sol_trajectory_oracle} decomposes into $M$ independent convex trajectory subproblems that can be solved in parallel. With \(\widehat{\bm X}^r\) obtained from \eqref{prob:sol_trajectory_oracle}, define the search direction and the corresponding first-order descent measure as
\begin{equation}
    \bm{D}_X^r\triangleq\widehat{\bm{X}}^r-\bm{X}^r,\quad
    \Delta_r\triangleq-\langle\bm{G}^r,\bm{D}_X^r\rangle\geq0.
    \label{eq:sol_trajectory_direction}
\end{equation}
For $\Delta_r>0$, an Armijo line search determines $\alpha_r\in\{1,\beta_{\rm A},\beta_{\rm A}^2,\ldots\}$ such that
\begin{equation}
    J(\bm{X}^r+\alpha_r\bm{D}_X^r)
    \leq
    J(\bm{X}^r)-\eta_{\rm A}\alpha_r\Delta_r,
    \label{eq:sol_armijo}
    \end{equation}
    where $\beta_{\rm A},\eta_{\rm A}\in(0,1)$. The trajectory is then updated as
    \begin{equation}
    \bm{X}^{r+1}=\bm{X}^r+\alpha_r\bm{D}_X^r.
    \label{eq:sol_trajectory_update}
\end{equation}
The corresponding physical boresights are recovered through \eqref{eq:sol_cap_parameterization}. Since $\bm{X}^r$ and $\widehat{\bm{X}}^r$ are both feasible for \eqref{prob:sol_trajectory_oracle}, any line-search point with $\alpha_r\in[0,1]$ remains feasible and therefore satisfies the original steering-range and slew constraints.

\addtolength{\topmargin}{0.07in}

\subsection{Overall Algorithm and Convergence Analysis}
\label{subsec:sol_algorithm}

\begin{algorithm}[t]
\caption{Overall Alternating Optimization Algorithm}
\label{alg:sol_overall}
\begin{algorithmic}[1]
\Require Feasible $(\mathcal{F}^0,\mathcal{C}^0)$.
\State Initialize $(\mathcal{W}^0,\boldsymbol{\nu}^0)$ by \eqref{eq:sol_w_update} and \eqref{eq:sol_nu_update}; set $r=0$.
\Repeat
    \State Update $\mathcal{C}^{r+1}$ via \eqref{eq:sol_polar_singular}-\eqref{eq:sol_polar_secular}.
    \State Compute $\bm{G}^r$ and solve \eqref{prob:sol_trajectory_oracle} to obtain $\widehat{\bm{X}}^r$.
    \State Form $\bm{D}_X^r$ and $\Delta_r$ by \eqref{eq:sol_trajectory_direction}.
    \If{$\Delta_r>0$}
        \State Determine $\alpha_r$ from \eqref{eq:sol_armijo} and update $\bm{X}^{r+1}$ by \eqref{eq:sol_trajectory_update}.
    \Else
        \State Set $\bm{X}^{r+1}=\bm{X}^r$.
    \EndIf
    \State Recover $\mathcal{F}^{r+1}$ from \eqref{eq:sol_cap_parameterization}.
    \State Update $(\mathcal{W}^{r+1},\boldsymbol{\nu}^{r+1})$ by \eqref{eq:sol_w_update} and \eqref{eq:sol_nu_update}.
    \State Set $r\leftarrow r+1$.
\Until{convergence}
\end{algorithmic}
\end{algorithm}

The overall algorithm is summarized in Algorithm~\ref{alg:sol_overall}.
Each block update does not increase the WMMSE objective. Specifically, each polarforming subproblem is globally solved, the boresight update ensures feasible descent through the Armijo line search, and the receive filters and MSE weights are optimally updated in closed form. Since the WMMSE objective is bounded below, the resulting objective sequence converges.

\section{Simulation Results and Discussion}

We consider the uplink transmission from $K=12$ GTs to a LEO satellite at an altitude of $600$ km, carrier frequency of $2$ GHz, and allocated bandwidth of $180$ kHz. The GTs are grouped into $G=3$ geographical clusters with $4$ GTs per cluster, a cluster radius of $20$ km, and an inter-cluster center spacing of $120$ km. The satellite employs a $4\times4$ UPA with half-wavelength element spacing, and the cosine-power directional model adopts $p=4$. The maximum boresight steering angle is $\theta_{\max}=50^\circ$, while the maximum angular speed is $\omega_{\max}=5^\circ/{\rm s}$ with a guard interval of $T_{\rm g}=2$ s, yielding a maximum inter-epoch slew of $\Omega_{\rm g}=10^\circ$. Unless otherwise stated, the GT transmit power is $23$ dBm and the receiver noise power is $\sigma^2=-118.42$ dBm. The GT polarization angles are independently drawn from $\mathcal{U}[0,\pi)$ and remain fixed over the observation interval. We consider the following schemes for comparison.  \textbf{\emph{Joint Design}}: both RA boresights and polarformers are optimized. \textbf{\emph{RA Only}}: boresights are optimized with fixed circular polarforming $\bm{c}_m[n]=[1,j]^T/\sqrt{2}$. \textbf{\emph{Polarforming Only}}: polarformers are optimized with fixed nadir-pointing boresights $\bm{f}_m[\ell]=\bm{e}_z$. \textbf{\emph{Benchmark}}: both are fixed as above.

\begin{figure}[t]
    \centering
    \includegraphics[width=0.8\columnwidth]{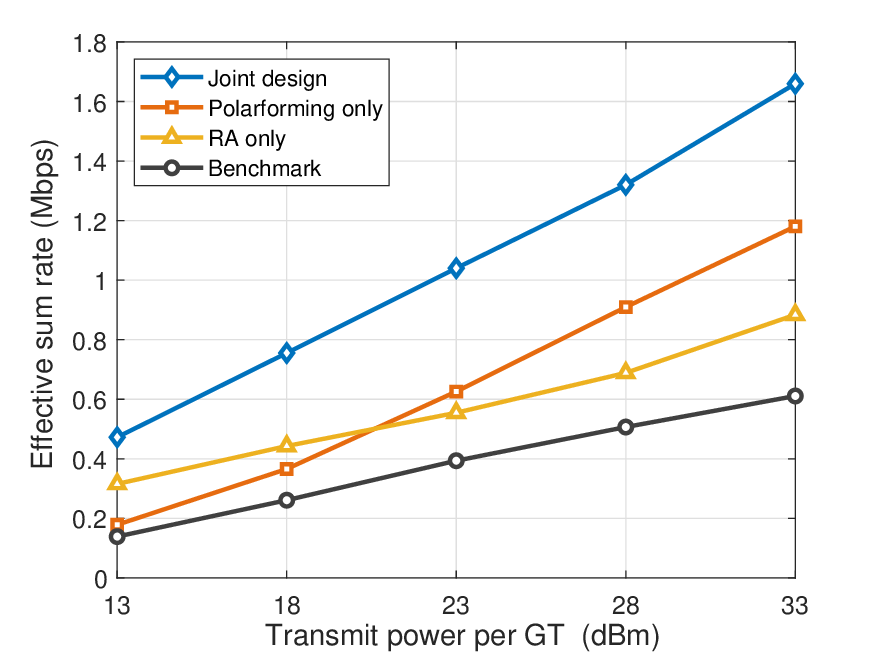}
    \caption{Effective sum rate versus GT transmit power.}
    \label{fig:power_range}
\end{figure}

Fig.~\ref{fig:power_range} shows the effective sum rate versus GT transmit power. At low transmit powers, RA steering is more beneficial in the noise-limited regime, where adjusting the antenna orientations enables more effective directional-gain allocation across different user clusters, thereby improving the received signal strength. As transmit power increases, multiuser interference becomes more prominent. In this regime, polarforming becomes increasingly effective by exploiting polarization diversity to improve the separability of spatially correlated user channels, eventually outperforming RA-only optimization. The joint design consistently achieves the highest throughput, reaching about $1.04$ Mbps at $23$ dBm, with gains of $67.7\%$ and $85.7\%$ over polarforming only and RA only, respectively. These results highlight the complementary roles of directional-gain allocation and polarization-domain channel shaping.

\begin{figure}[t]
    \centering
    \includegraphics[width=0.8\columnwidth]{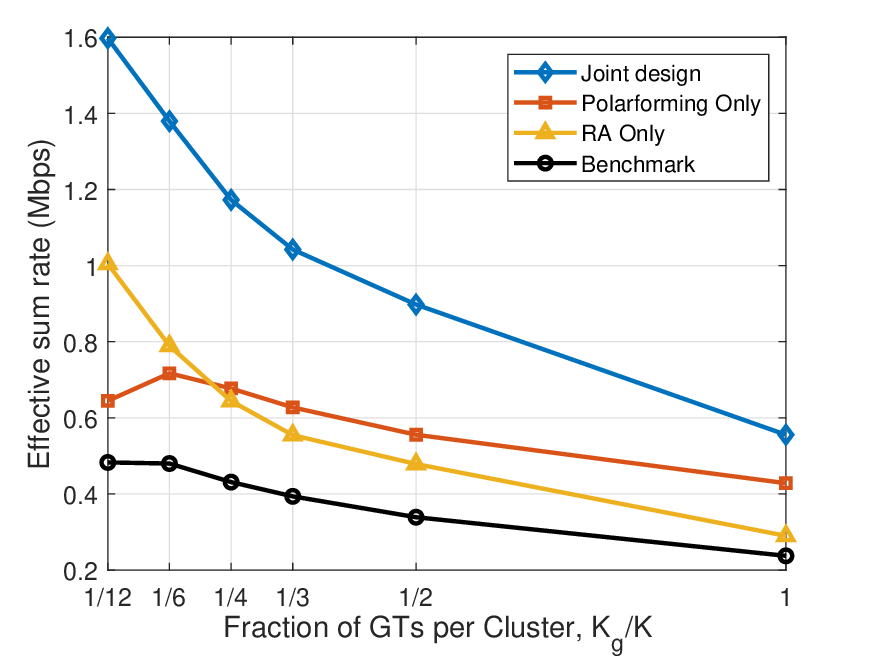}
    \caption{Effective sum rate versus the fraction of GTs per cluster.}
    \label{fig:gt_fraction}
\end{figure}

Fig.~\ref{fig:gt_fraction} evaluates the impact of user clustering while fixing the total number of GTs at $K=12$. We vary the number of clusters $G$ and the number of GTs per cluster $K_g$ while maintaining $GK_g=12$, such that a larger $K_g/K$ corresponds to fewer but more densely populated clusters. As $K_g/K$ increases, the effective sum rate generally decreases because the users become more spatially concentrated and hence more difficult to separate in the spatial domain. In particular, RA-only optimization outperforms polarforming-only optimization when users are distributed across many clusters, whereas polarforming becomes relatively more effective as users become concentrated within fewer clusters. This trend indicates that RA mainly exploits spatial and directional diversity across geographically separated users, while polarforming provides an additional degree of freedom for separating users with similar spatial directions. The joint design consistently achieves the highest throughput, further confirming the benefits of combining these two capabilities.

\section{Conclusion}

This paper studied joint antenna rotation and polarforming for satellite communications. We developed a WMMSE-based alternating algorithm with semi-closed-form polarforming updates and feasible RA-trajectory optimization via exact boresight parameterization. Simulations demonstrated substantial gains over benchmark schemes. Antenna rotation redistributes directional gains across user clusters and is particularly beneficial at low transmit powers, whereas polarforming improves channel separability and becomes increasingly valuable under stronger multiuser interference. These findings demonstrate the complementary benefits of directional and polarization-domain channel shaping in satellite communications.

\end{document}